\documentclass{pnastwo}

\usepackage[dvips]{graphicx}

\usepackage{amssymb,amsfonts,amsmath}

\contributor{Proceedings
of the National Academy of Sciences of the United States of America, in press}
\url{www.pnas.org/cgi/doi/10.1073/pnas.0913904107}
\begin{document}

%%%%%%%%%%%%%%%%%%%%%%%%%%%%%%

%% For titles, only capitalize the first letter
%% \title{Almost sharp fronts for the surface quasi-geostrophic equation}

\title{AGN feedback in clusters of galaxies}

%% Enter authors via the \author command.  
%% Use \affil to define affiliations.
%% (Leave no spaces between author name and \affil command)
%% Note that the \thanks{} command has been disabled in favor of
%% a generic, reserved space for PNAS publication footnotes.
%% One number for each institution.
%% The same number should be used for authors that
%% are affiliated with the same institution, after the first time
%% only the number is needed, ie, \affil{number}{text}, \affil{number}{}
%% Then, before last author ...
%% \and
%% \author{<author name>
%% \affil{<number>}{}}
%% For example, assuming Garcia and Sonnery are both affiliated with
%% Universidad de Murcia:
%% \author{Roberta Graff\affil{1}{University of Cambridge, Cambridge,
%% United Kingdom},
%% Javier de Ruiz Garcia\affil{2}{Universidad de Murcia, Bioquimica y Biologia
%% Molecular, Murcia, Spain}, \and Franklin Sonnery\affil{2}{}}

\author{Elizabeth L. Blanton\affil{1}{Institute for Astrophysical Research and Astronomy Department, Boston University, 
725 Commonwealth Avenue, Boston, MA 02215, USA},
T. E. Clarke\affil{2}{Naval Research Laboratory, 455 Overlook Avenue SW, Washington, DC 20375, USA},
 Craig L. Sarazin\affil{3}{Department of
Astronomy, University of Virginia, P. O. Box 400325, Charlottesville, VA 22904-4325, USA}, 
Scott W. Randall\affil{4}{Harvard-Smithsonian Center for Astrophysics, 60 Garden Street, Cambridge, MA 02138, USA},
and {Brian R. McNamara\affil{4}{}\affil{5}{Department of Physics and Astronomy, University of Waterloo, Waterloo, ON N2L 2G1,
Canada; Perimeter Institute for Theoretical Physics, 31 Caroline Street, N. Waterloo, Ontario N2L 2Y5, Canada}}}

\contributor{Proceedings of the National Academy of Sciences
of the United States of America, in press}

%% The \maketitle command is necessary to build the title page.
\maketitle

%%%%%%%%%%%%%%%%%%%%%%%%%%%%%%%%%%%%%%%%%%%%%%%%%%%%%%%%%%%%%%%%
\begin{article}

\begin{abstract}
Observations made during the last ten years with the Chandra X-ray Observatory have shed much light on the cooling gas
in the centers of clusters of galaxies and the role of active galactic nucleus (AGN) heating.  Cooling of the hot 
intracluster
medium in cluster centers can feed the supermassive black holes found in the nuclei of the dominant cluster
galaxies leading to AGN outbursts which can reheat the gas, suppressing cooling and large amounts of star 
formation.  AGN heating can come in the form of shocks, buoyantly rising bubbles that have been inflated 
by radio lobes, and the dissipation of sound waves.
\end{abstract}

%% When adding keywords, separate each term with a straight line: |

%% Optional for entering abbreviations, separate the abbreviation from
%% its definition with a comma, separate each pair with a semicolon:
%% for example:
%% \abbreviations{SAM, self-assembled monolayer; OTS,
%% octadecyltrichlorosilane}

% \abbreviations{}

%% The first letter of the article should be drop cap: \dropcap{}
%\dropcap{I}n this article we study the evolution of ''almost-sharp'' fronts

%% Enter the text of your article beginning here and ending before
%% \begin{acknowledgements}
%% Section head commands for your reference:
%% \section{}
%% \subsection{}
%% \subsubsection{}

\section{Introduction}
\dropcap{C}handra observations of cavities in cluster centers in the last decade have brought the issue of 
feedback to the center stage in astrophysics. Feedback from AGN driven by supermassive black holes in the cores of massive galaxies is now seen as
a necessary ingredient to adequately describe the formation and evolution of galaxies and the clusters in which
they sometimes reside.  The energy injected by AGN is required to produce a galaxy luminosity function in simulations
(e.g. \cite{croton06}) that matches observations.  This feedback may set the upper limit to the observed masses
of galaxies and contributes to cluster preheating and the observed ``entropy floors'' seen in cluster centers.
It may affect scaling relations, such as the relationship between X-ray luminosity and temperature and potentially
affects cluster properties that are used for constraining cosmological models, such as the gas mass fraction
(e.g. \cite{allen04}).  Data from the {\it{Chandra X-ray Observatory}} over the last ten years reveal AGN at work in
the centers of clusters, inflating bubbles that rise buoyantly through the intracluster medium (ICM), sometimes
producing shocks and sound waves.

Many of the most spectacular {\it{Chandra}} images of AGN feedback are of interactions in the centers of 
cooling flow (sometimes referred to as ``cool core'') clusters of galaxies.  Under the assumption of hydrostatic
equilibrium, the X-ray-emitting gas in a cluster is the most dense at the cluster center.  The gas will cool
as it radiates, and the cooling time is inversely proportional to the density of the gas.  Therefore, gas will
cool at cluster centers before it cools at larger radii.  If the cooling time is shorter than the cluster age, the
central gas will cool, lose pressure, and outer gas must flow in to maintain hydrostatic equilibrium (see 
\cite{fabian94} for a review).
Data from earlier X-ray observatories such as {\it{Einstein}} \cite{white97}, the ROentgen SATellite ({\it{ROSAT}}) 
\cite{peres98},
and the Advanced Satellite for Cosmology and Astrophysics ({\it{ASCA}}) \cite{white00} were used to
give estimates of the gas inflow rates of typically hundreds or thousands of solar masses per year.  The classic
``cooling flow problem'' has been that sufficient quantities of cool gas and/or star formation were not detected
to match the predictions from the X-ray data.

{\it{Chandra}} and {\it{XMM-Newton}} observations have shown that while the gas temperature drops in
the centers of cooling flow clusters, the temperature reaches a threshold value below which there is little cooling.
This is seen in the temperature profiles as well as high resolution spectroscopy.  An important early
result from {\it{XMM-Newton}} high resolution spectra from cooling flows was that the emission lines
from cool gas were not present at the expected levels, and the spectra were well-fitted by a cooling flow model with a low temperature
cutoff \cite{peterson01,peterson03}.  This cutoff is typically one-half to one-third of the average cluster temperature,
and is usually not much below 1 keV.  The gas is either then cooling non-radiatively or being heated to higher 
temperatures.
Throughout this paper, we use the term ``cooling flow'' to indicate clusters with central drops in temperature and
large central density peaks above that of a typical $\beta$-model.

There is a clear association with the presence of a cooling flow in a cluster and the central galaxy hosting a radio
source.
Using {\it{Einstein}} data, \cite{burns90} found $\approx70\%$ of cooling flow clusters had giant elliptical galaxies that
emitted in the radio, as compared to only $\approx20\%$ of central galaxies in non-cooling flows.  Recent studies,
depending on the sample, find a fraction of up to 100\% for cooling flows and 45\% for non-cooling flows
\cite{mittal09}
using {\it{Chandra}} data.  This association is consistent with the cooling gas supplying fuel to the
central black hole, leading to the AGN outburst observed in the radio.  As seen in {\it{Chandra}} images, the radio jets and lobes 
associated with the AGN significantly affect the X-ray gas, inflating ``cavities'' or ``bubbles'' in the cluster
centers.  Clear correspondence between regions of radio emission and deficits in the X-ray are seen in a number of
cases (e.g. Hydra A \cite{mcnamara00,wise07}, Perseus \cite{fabian00}, 
Abell 2052, \cite{blanton01}, Abell 2597 \cite{mcnamara01,clarke05}, Abell 2199 \cite{johnstone02},
Centaurus \cite{fabian05}, Abell 4059 \cite{heinz02,choi04},
Abell 262 \cite{blanton04,clarke09}, MS0735.6+7321 \cite{mcnamara05},
M87/Virgo \cite{forman07}).  
The X-ray gas, in turn, affects the radio lobes by confining and distorting them.  Energy
can be transported to the ICM through bubbles that rise buoyantly to larger cluster radii, as well as through
weak shocks and sound waves generated by AGN outbursts.  The energy transferred to the cluster gas can diminish
the cooling, giving an explanation for the lack of copious quantities of cool gas and star formation seen in the
centers of cooling flow clusters.
Composite images of A2052 and A262 are shown in Figs.\ 1 and 2, respectively, where blue is
the X-ray emission as seen with {\it Chandra}, red is optical emission, and yellow/green (in the cluster centers)
is 1.4 GHz radio emission.

\section{Cavity properties}

The sizes of the cavities or bubbles created by the radio lobes evacuating regions of the ICM vary widely from
a few kpc (e.g. Abell 262 \cite{blanton04,clarke09}) to hundreds of kpc (e.g. MS0735.6+7321
\cite{mcnamara05}) in diameter.  The cavities are often surrounded by bright rims or shells, where the 
gas from the cavity
interiors has been compressed.  In addition to the visual impression that the radio lobes are removing the 
X-ray gas from the cavities, we can quantitatively confirm this by measuring the mass of gas in the shells and
comparing it to the mass of gas that should fill the central region of the cluster estimated by extrapolating
the density profile of the gas at outer radii inward.  In Abell 2052 \cite{blanton01,blanton03}, for instance,
the mass in the shells compressed by the two radio lobes is approximately $5\times10^{10}~M_{\odot}$, consistent
with the mass predicted to have filled the cavities, based on the extrapolation of the density profile.

The X-ray gas in the compressed shells surrounding the radio lobes is generally found to be cool, and does not
show evidence of strong shock heating as had been predicted with early models of radio source / ICM interaction
(e.g. \cite{heinz98}).  Cool gas is seen surrounding the inner radio lobes of, for example, 
Hydra A \cite{nulsen02}, Perseus A \cite{schmidt02}, 3C 317 in Abell 2052 \cite{blanton01,blanton03},
and B2 0149+35 in Abell 262 \cite{blanton04}.
The cooling time within the shells is typically longer than the radio source age, suggesting that the gas in the
shells had to cool closer in to the cluster center before being pushed out by the radio lobes.  In the case of
Abell 2052, the shell cooling time is approximately $3\times10^8$ yr while the age of the radio source is 
approximately $10^7$ yr.
The bright X-ray shells are sometimes coincident with optical line emission, such as $H\alpha$, representing
gas that has cooled to temperatures of approximately $10^4$ K \cite{blanton01,blanton04,fabian03b}.  Therefore, while most of the cluster
gas cools to only one-third to one-half of the average ICM temperature, at least some quantity of gas can cool
to much lower temperatures and sometimes results in star formation \cite{mcnamara97}.

Assuming some geometry for the shells surrounding the radio lobes and fitting the spectra from these regions,
the densities and pressures can be determined.  
Alternatively, the pressure in the X-ray gas is determined from undisturbed regions in the cluster
at the same radii as the cavities, yielding similar results \cite{mcnamara07}.
The pressures found are often higher (by approximately an
order of magnitude) than the pressures derived from radio observations, assuming equipartition, for the cavities.  
The cavities would be expected to collapse quickly if they were indeed underpressured,
and it is therefore assumed that some additional source of pressure must be filling the cavities.
The origin of this additional pressure is still unknown, but candidates are low energy, relativistic electrons,
a higher ratio of ions to electrons than is typically assumed, a magnetic field with a value higher than
the equipartition value, or very hot, diffuse, thermal gas.  Hot, diffuse, thermal gas has not been detected
filling the radio cavities, but limits have been placed in a few cases ($>15$ keV in Hydra A \cite{nulsen02},
$>11$ keV in Perseus \cite{schmidt02}, and $>20$ keV in Abell 2052 \cite{blanton03}).

The cavities are eventually expected to break up due to Rayleigh-Taylor and
Kelvin-Helmholtz instabilities.  Rayleigh-Taylor instabilities
are a possible explanation for the spur of gas jutting into the N cavity in Abell
2052 \cite{blanton01,blanton03,blanton09}.  The cavities can be maintained against these instabilities by
magnetic draping \cite{lyutikov06} or pressure related to bubble inflation \cite{pizzolato06}.

\section{Buoyant bubbles}

Since the density inside the radio cavities is much lower than that of the ambient gas, the cavities should be 
buoyant, and will rise out to larger cluster radii.  Such outer cavities are called ``ghost cavities'' or 
``ghost bubbles'' because they are thought to be the result of earlier AGN outbursts.  In addition, if these
cavities are filled with radio lobes, the lobes often have steep spectra,
and radio emission at the commonly-observed radio 
frequency of 1.4 GHz is sometimes faint or absent.  Radio
observations at lower frequencies have shown that the radio sources extend into these outer cavities in
many cases.

Examples of systems with outer cavities are Perseus \cite{fabian00}, Abell 2597 \cite{mcnamara01,clarke05},
Abell 262 \cite{clarke09}, Abell 2052 \cite{blanton09}, and Hydra A \cite{wise07}.  In all of these
cases, when the cavities seemed devoid of radio emission at higher frequencies, lower frequency radio emission was
found to extend farther from the cluster center to fill the outer cavities.

\subsection{Entrainment of cool gas}

Cool gas can be entrained along with radio lobes as they propagate into the ICM.
There is a clear association with arcs of cool gas and the radio
jets and lobes in the M87/Virgo system \cite{forman07,forman05,young02}.
This is also seen in the center of Abell 133, where a cool X-ray plume appears related to a
detached radio lobe or relic \cite{fujita02}.  In the Perseus cluster 
\cite{fabian03b},
plumes of cool, $\approx10^4$ K gas seen in $H\alpha$ appear to have risen away from the cluster center, having
been dragged by one of the outer cavities.
%In addition to there being evidence in the temperature structure of the gas to support the scenario
%of entrainment, 
The metallicity structure
is also consistent with the entrainment scenario \cite{kirkpatrick09}.  Analysis of deep {\it Chandra} observations of the Hydra A cluster reveals higher
metallicity gas along the radio jets and lobes, indicating that the gas is being redistributed from
the center (where the gas metallicity is typically higher) to larger radii in the cluster \cite{kirkpatrick09}.

\subsection{X-ray cavities as radio calorimeters}

Rather than relying on radio observations to estimate the energy input from the radio sources where the ratio
of radiative power to kinetic power can vary widely, we can instead use
observations of the effects that the radio sources have on the X-ray-emitting ICM to determine how much
energy is being deposited into the cluster gas.
As in \cite{mcnamara00} and \cite{churazov02}, the energy input is a combination of the internal energy in the cavity
and the work done in inflating the cavity, or the enthalpy:

\begin{equation}
\frac{1}{(\gamma-1)}PV + PdV = \frac{\gamma}{(\gamma-1)}PV
\end{equation}
where $P$ is the pressure measured in the X-ray shells (which are assumed to be in pressure equilibrium with
the cavities), and $V$ is the volume of the cavities.  The value of $\gamma$ is the mean adiabatic index
of the gas filling the cavities and is 4/3 for relativistic gas or 5/3 for non-relativistic, monatomic gas.  In the
relativisitic case, the total energy input is then $4PV$, and in the non-relativistic case it is $(5/2)PV$.
To estimate an average energy input rate, we need the radio source outburst rate.  In some
cases, this can be measured directly from the X-ray data, when there are multiple sets of bubbles.  Taking the
projected distances between outer cavities and inner cavities and calculating the time to rise buoyantly to
the outer position, the time between outbursts can be estimated.  The time between outbursts is typically
a few tens of millions of years (e.g. Perseus \cite{fabian00}; A262 \cite{clarke09}, A2052 \cite{blanton09}).
In cases without outer cavities, an approximate value based on other systems is assumed.

\subsection{Does energy input from bouyantly rising bubbles offset cooling?}

We can compare the energy injection rate from AGN-inflated cavities that rise buoyantly in clusters
to the luminosity of cooling gas in the cluster centers.  This cooling luminosity is gvien by

\begin{equation}
L_{cool}=\frac{5}{2}\frac{\dot{M}}{\mu m_p}kT
\end{equation}
where $\dot{M}$ is the mass deposition rate into the cluster center of the cooling gas, T is the 
upper temperature from which the gas is cooling, and $\mu$ is the mean molecular weight with
$\mu m_p$ being the average mass per particle in the gas.
In many cases, the energy input rate from the buoyant bubbles is enough to offset the cooling
gas (e.g. Abell 2052 \cite{blanton01,blanton03,blanton09} with a rate of $\approx3\times10^{43}$ erg
s$^{-1}$ and Hydra A \cite{mcnamara00,nulsen02,david01} with a rate 
of $\approx3\times10^{44}$ erg s$^{-1}$).  In some cases, the bubbles did not, at least initially, appear
to supply enough energy to the ICM (e.g. Abell 262 \cite{blanton04}, where the energy input
rate is approximately an order of magnitude too low).

In addition to detailed studies of individual objects, studies of samples of cooling flow clusters
with central AGN have been undertaken \cite{birzan04,birzan08,diehl08,dunn06,rafferty06}.
Thirty-three systems were included in a sample in \cite{rafferty06}, and it was found that,
on average, $4PV$ of enthalpy per cavity supplies enough energy to substantially suppress cooling in
clusters. 

\section{Shocks}

Simulations of AGN outbursts in cluster centers show the creation of shocks as well as cavities
related to the radio lobes (e.g. \cite{heinz06}).  These ``cocoon'' shocks can heat the cluster
gas not only in the direction parallel to the propagation of the radio lobes, but perpendicular to
their propagation axis, as well.  Therefore, energy can be distributed throughout the cluster
center if shocks are present.

Until recently, shocks were not seen associated with AGN radio lobes in cluster cooling flows, but
in the last few years, several have been detected.
A spectacular example is the cluster MS0735.6+7321 \cite{mcnamara05} which hosts extremely large
cavities with diameters of approximately 200 kpc.  The total energy injection required to inflate
the cavities and produce the observed shocks is $6\times10^{61}$ erg, making this the most powerful
radio outburst known.
The shock energy input rate is given by

\begin{equation}
\Pi_{s}=\frac{(\gamma+1)P}{12\gamma^2}\left(\frac{\omega}{2\pi}\right)\left(\frac{\delta P}{P}\right)^3
\end{equation}
where $P$ is the pre-shock pressure, $\gamma=5/3$, and $2\pi/\omega$ is the time interval between
shocks \cite{mcnamara07}.
A deep {\it Chandra} image of M87/Virgo revealed an approximately spherical shock in the cluster
center \cite{forman07}.  The shock is most clearly seen in a hard-band image, and is fairly
weak with a Mach number of 1.2.
A Mach 1.65 shock was found in the Hercules cluster \cite{nulsen05}, with the total
energy deposited $3\times10^{61}$ erg.  There is a weak shock (Mach 1.3) outside of one of the bubble
rims in the center of the Perseus cluster \cite{graham08}.  As is common, the density jump measured is consistent with a 
shock model, but a rise in temperature is not measured.  Two shocks are seen surrounding the center 
of Abell 2052 \cite{blanton09}, both with Mach 1.2.  The temperatures across the shocks are
consistent, within the errors, to the jumps expected in the shock models, but the best-fitting
temperature values are constant across the fronts.
If an increase in temperature is not measured behind a shock, \cite{fabian06} has suggested that
shocks may be isothermal if conduction is efficient in these regions.  However, the temperature
rises are very difficult to detect given the narrow width of the shock fronts, adiabatic expansion, and the 
difficulty in modeling the large
amount of cluster gas at larger radii projected onto these regions when fitting the X-ray spectra.
Based on further analysis of the Perseus system, \cite{graham08} conclude that mixing of 
post-shock gas with cool gas associated with optical-line emission may result in the lack of
temperature rise seen associated with the shock.

\section{Sound waves}

In addition to weak shocks in the centers of cooling flow clusters, sound or pressure waves have
also been observed.  These features are most prominent in the Perseus cluster \cite{fabian06,fabian03a}.
They appear as $5-10\%$ increases in pressure in ripple-like patterns surrounding
cluster centers.  Changes in temperature are not seen in these regions, and it is likely that
their energy is dissipated by viscosity.  There may be one or more set of ripples corresponding
to each AGN outburst depending on variations in energy output that can occur during a single
outburst.  Assuming they are associated with single outbursts, the ripple separations in Perseus
imply a repetition rate of $10^7$ yr \cite{fabian03a}.  
As shocks
propagate outward from cluster centers, they weaken and may appear as sound waves.  In addition
to the Perseus cluster, sound wave ripples are seen in the Centaurus cluster \cite{sanders08} and 
M87/Virgo \cite{forman07}.  Concentric ripple-like features are also seen surrounding
the center of Abell 2052, but current analysis shows that these are consistent with weak shocks
\cite{blanton09}.

A key difference between shock and sound wave heating is that sound wave dissipation depends
on transport coefficients and shock dissipation does not \cite{mcnamara07}.  
The sound wave dissipation rate is sensitive to temperature and would be less effective in
lower temperature systems.  This may be offset by a higher frequency of outbursts in cooler
clusters \cite{mcnamara07}.  
One of the important implications of the sound waves is that energy can be deposited over $4\pi$ steradians, 
and not only along the narrow angle of the radio jets.

\section{Case studies:  Abell 2052 and Abell 262}

Abell 2052 is one of the few clusters that showed evidence for radio bubbles using data from an
X-ray telescope prior to the {\it Chandra} / {\it XMM-Newton} era ({\it ROSAT} \cite{rizza00}).
It was first observed with {\it Chandra} for 37 ksec in Cycle 1 in 2000, and subsequently observed
for 129 ksec in Cycle 6 in 2006 and 497 ksec in Cycle 10 in 2009.  A three-color image from
the total 657 ksec cleaned dataset is shown in Fig.\ 3, where the colors represent different
energy ranges (red $0.3-1.0$ keV, green $1.0-2.0$ keV, and blue $2.0-10.0$ keV).
Clear bubbles are seen to the north and south of the cluster center, and a filament extends
into the northern bubble.  Surrounding the bubbles are bright shells of emission, and outside of
these shells a jump is the surface brightness is visible extending around the cluster center.
A second jump is seen outside this one, and it appears sharper to the NE than to the SW.  It is
possible that this feature is a cold front rather than a shock.

An unsharp-masked image for the total 657 ksec dataset is shown in Fig.\ 4  in grayscale with 1.4 GHz
radio emission from the Very Large Array Faint Images of the Radio Sky at Twenty-cm (VLA FIRST) 
survey \cite{becker05} superposed in green contours.  Concentric ripple-like
features are seen surrounding the cluster center, and these are modeled as shocks in
\cite{blanton09} based on the earlier 163 ksec dataset.  These features are at 31 and 46 kpc
from the AGN and the density jumps associated with them are both consistent with shocks with Mach
number 1.2.  
The separation of the potential shock features is consistent with a radio source outburst
rate of $2\times10^7$ yr.  This same cycle time is found when considering the locations of outer
cavities in the system.  A small cavity appears bounded by a narrow filament to the NW of the
N bubble, and the S cavity appears split into two cavities, with the outer one to the SE.
Assuming the outer cavities rose buoyantly to their current positions, we find a cycle time
of $2-4 \times10^7$ yr depending on whether they rose at 0.5 times or 1 time the sound speed.

In \cite{blanton09}, we found a mass deposition rate of $\dot{M} = 55\pm4~~M_{\odot}$ yr$^{-1}$
and a cooling luminosity of $L_{cool}=5.4\times10^{43}$ erg s$^{-1}$.
We find the rate of energy input from the buoyantly rising bubbles assuming they are filled
with relativistic plasma ($\gamma=4/3$) to be $3-6\times10^{43}$ erg s$^{-1}$ depending on the 
buoyancy rise time.  The input from shock heating is only $1\times10^{43}$ erg s$^{-1}$, however,
the combination of heating from shocks and buoyantly rising bubbles can offset the cooling in this
system.

Abell 262 was first observed by {\it Chandra} in Cycle 2 in 2001 for 30 ksec.  It was subsequently observed
in Cycle 8 in 2006 for 112 ksec.  A mass deposition rate of  $\dot{M} = 19^{+6}_{-5}~~M_{\odot}$ yr$^{-1}$
and a cooling luminosity of $L_{cool}=1.3\times10^{43}$ erg s$^{-1}$ were found in \cite{blanton04}.  Based
on the one clear bubble seen in the earlier dataset, it was found that the radio source energy input fell
more than an order of magnitude short of offsetting the cooling \cite{blanton04}, and it was concluded that the
AGN may have had a more powerful outburst in the past so that cooling could be offset on average.  In this study,
a repetition rate of $1\times10^8$ yr was assumed for the AGN outbursts.

Analyzing the longer dataset, however, we found in \cite{clarke09} that the radio source can offset the cooling.
This analysis was based on both a deep {\it Chandra} observation and low-frequency radio observations.  Several
more bubbles were found, and a tunnel made up of multiple bubbles rising with short time intervals was found to the
west of the cluster center.  Using the multiple sets of bubbles, the radio source comes within a factor of two of
quenching the cooling.  {\it Chandra} X-ray residual images superposed with radio emission at
1440 and 610 MHz are shown in Fig.\ 5.  
It may well be that further analysis of deeper multi-frequency observations of objects 
in samples such as \cite{rafferty06} that seem to have insufficient energy input rates would reveal 
that the AGN are indeed powerful enough to offset cooling.

%% == end of paper:

%% Optional Materials and Methods Section
%% The Materials and Methods section header will be added automatically.

%% Enter any subheads and the Materials and Methods text below.
%\begin{materials}
% Materials text
%\end{materials}

%% Optional Appendix or Appendices
%% \appendix Appendix text...
%% or, for appendix with title, use square brackets:
%% \appendix[Appendix Title]

\begin{acknowledgments}
We are very grateful to our many collaborators.  We thank the organizers for arranging a wonderful meeting 
highlighting the results from {\it Chandra} over the last ten years.  ELB was supported by the National Aeronautics
and Space Administration through {\it Chandra} award GO9-0147X.  CLS was supported in part by {\it Chandra} award
GO9-0035X.  Basic research in radio astronomy at the Naval Research Laboratory is supported by 6.1 base funding.
\end{acknowledgments}

%% PNAS does not support submission of supporting .tex files such as BibTeX.
%% Instead all references must be included in the article .tex document. 
%% If you currently use BibTeX, your bibliography is formed because the 
%% command \verb+\bibliography{}+ brings the <filename>.bbl file into your
%% .tex document. To conform to PNAS requirements, copy the reference listings
%% from your .bbl file and add them to the article .tex file, using the
%% bibliography environment described above.  

%%  Contact pnas@nas.edu if you need assistance with your
%%  bibliography.

% Sample bibliography item in PNAS format:
%% \bibitem{in-text reference} comma-separated author names up to 5,
%% for more than 5 authors use first author last name et al. (year published)
%% article title  {\it Journal Name} volume #: start page-end page.
%% ie,
% \bibitem{Neuhaus} Neuhaus J-M, Sitcher L, Meins F, Jr, Boller T (1991) 
% A short C-terminal sequence is necessary and sufficient for the
% targeting of chitinases to the plant vacuole. 
% {\it Proc Natl Acad Sci USA} 88:10362-10366.

%% Enter the largest bibliography number in the facing curly brackets
%% following \begin{thebibliography}

\end{article}
%%%%%%%%%%%%%%%%%%%%%%%%%%%%%%%%%%%%%%%%%%%%%%%%%%%%%%%%%%%%%%%%

%% Adding Figure and Table Referencess
%% Be sure to add figures and tables after \end{article}
%% and before \end{document}

%% For figures, put the caption below the illustration.
%%
%% \begin{figure}
%% \caption{Almost Sharp Front}\label{afoto}
%% \end{figure}

%% For Tables, put caption above table
%%
%% Table caption should start with a capital letter, continue with lower case
%% and not have a period at the end
%% Using @{\vrule height ?? depth ?? width0pt} in the tabular preamble will
%% keep that much space between every line in the table.

%% \begin{table}
%% \caption{Repeat length of longer allele by age of onset class}
%% \begin{tabular}{@{\vrule height 10.5pt depth4pt  width0pt}lrcccc}
%% table text
%% \end{tabular}
%% \end{table}

%% For two column figures and tables, use the following:

%% \begin{figure*}
%% \caption{Almost Sharp Front}\label{afoto}
%% \end{figure*}

%% \begin{table*}
%% \caption{Repeat length of longer allele by age of onset class}
%% \begin{tabular}{ccc}
%% table text
%% \end{tabular}
%% \end{table*}

\begin{figure}
\begin{center}
\includegraphics[width=.7\textwidth]{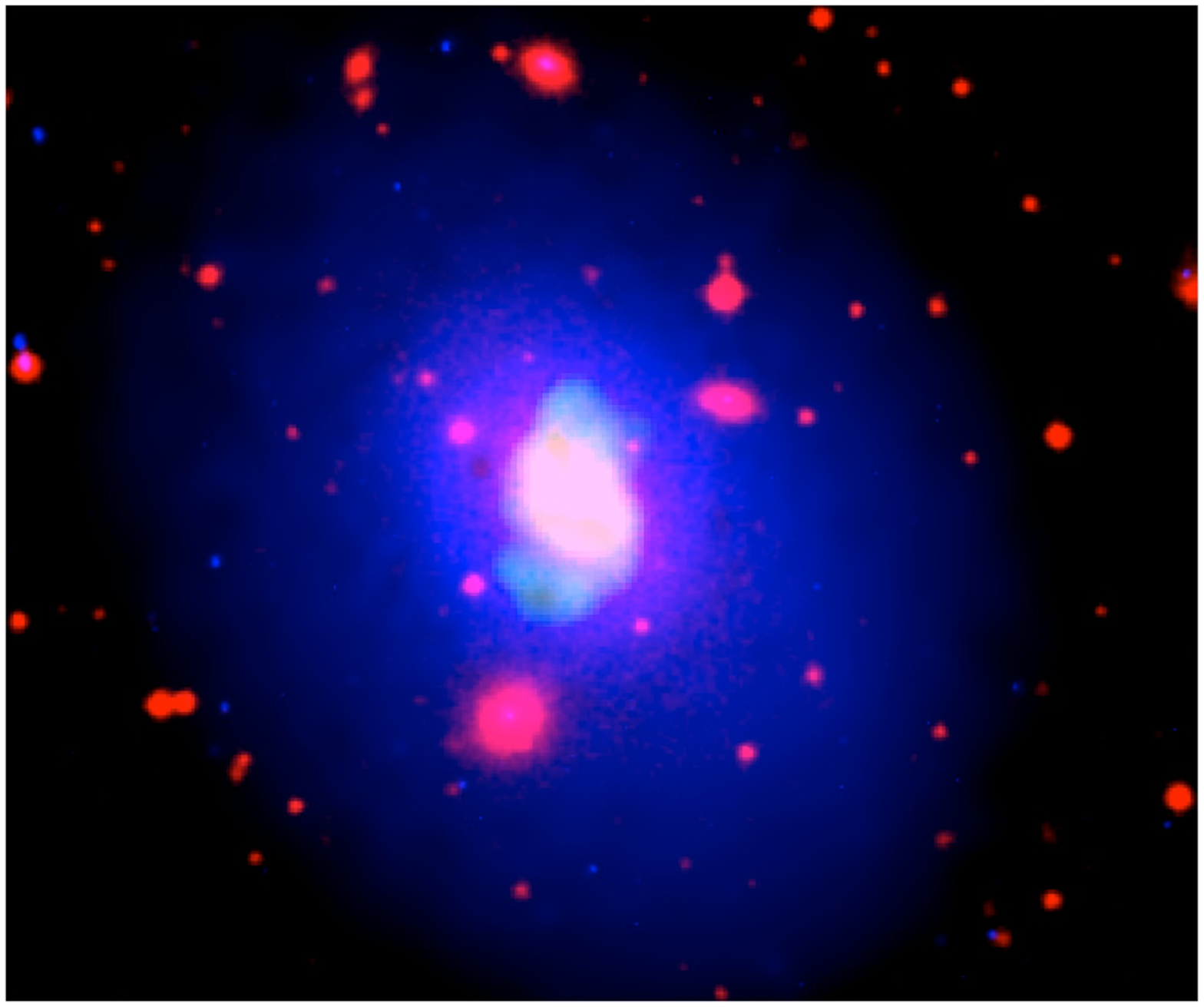}
\caption{Composite X-ray (blue, {\it Chandra} 657 ksec image), optical (red, Digitized Sky Survey), radio (green/yellow, 1.4 GHz \cite{becker05}) $5'.6\times4'.7$ ($230\times190$ kpc) image of Abell 2052.}
\end{center}
\end{figure}

\begin{figure}
\begin{center}
\includegraphics[width=.7\textwidth]{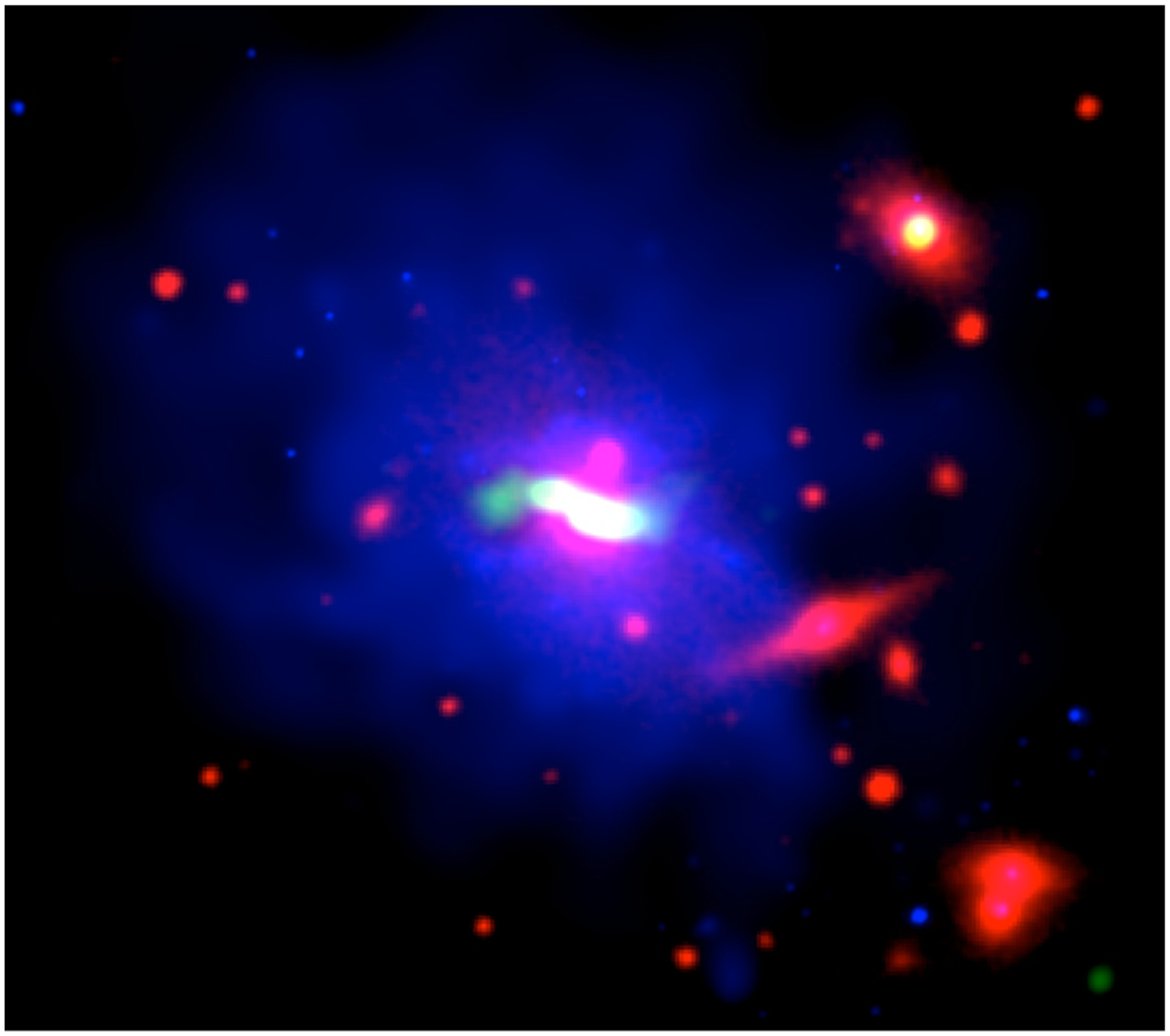}
\caption{Composite X-ray (blue, {\it Chandra} 139 ksec image), optical (red, Digitized Sky Survey), radio (green/yellow, 1.4 GHz \cite{clarke09}) $5'.6\times4'.7$ ($110\times90$ kpc) image of Abell 262.}
\end{center}
\end{figure}

\begin{figure}
\includegraphics[angle=-90,width=.6\textwidth]{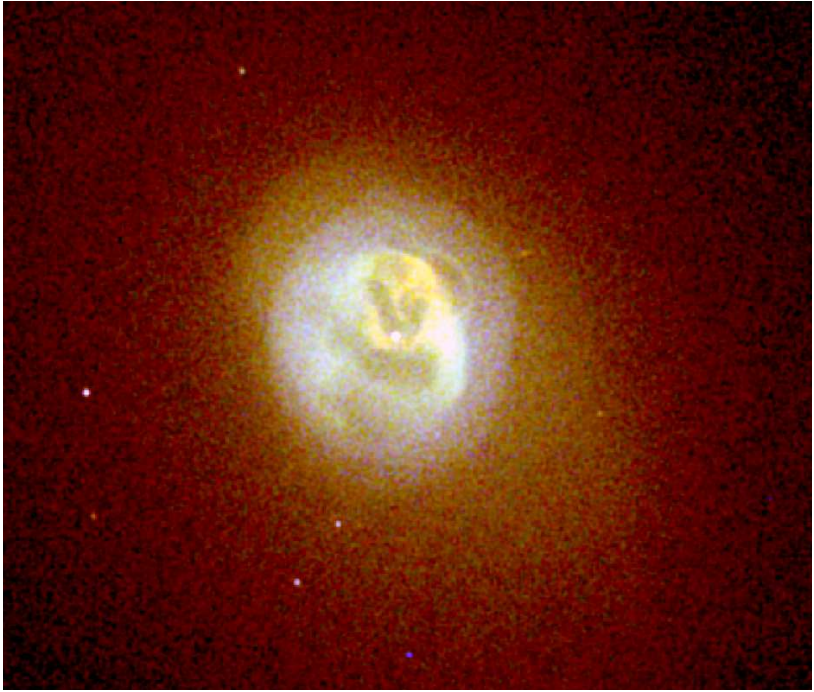}
\caption{Three-color {\it Chandra} 657 ksec $4'.3\times3'.7$ ($180\times150$ kpc) 
image of Abell 2052 (red = 0.3-1 keV, green = 1-2 keV, 
blue = 2-10 keV).}
\end{figure}

\begin{figure}
\includegraphics[width=.6\textwidth]{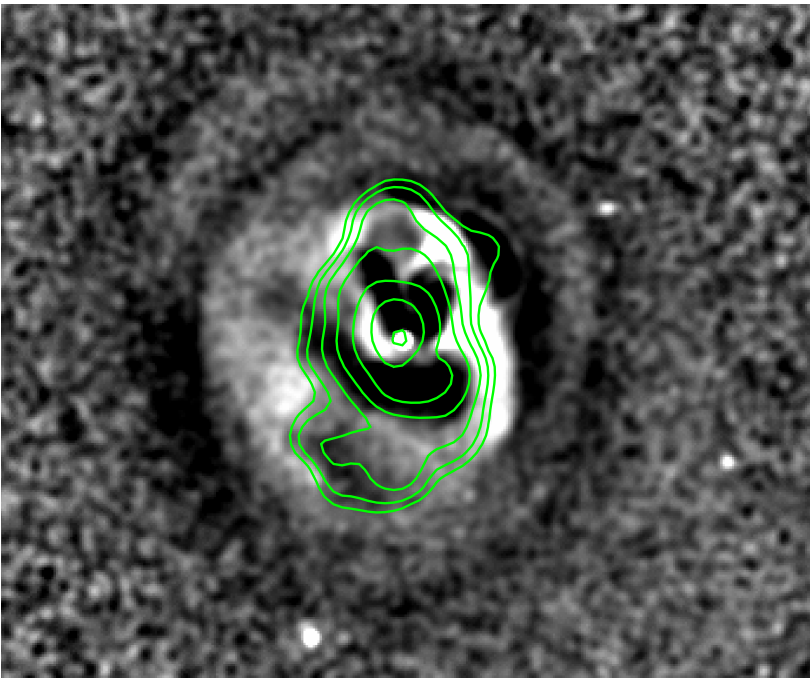}
\caption{Unsharp-masked 657 ksec {\it Chandra} $2'.8\times2'.4$ ($115\times95$ kpc) image of Abell 2052 with 1.4 GHz radio contours from the
VLA FIRST survey superposed.  Multiple sets of bubbles are seen, as well as ripple-like jumps in 
surface brightness that are consistent with weak shocks.}
\end{figure}

\begin{figure}
\includegraphics[width=.6\textwidth]{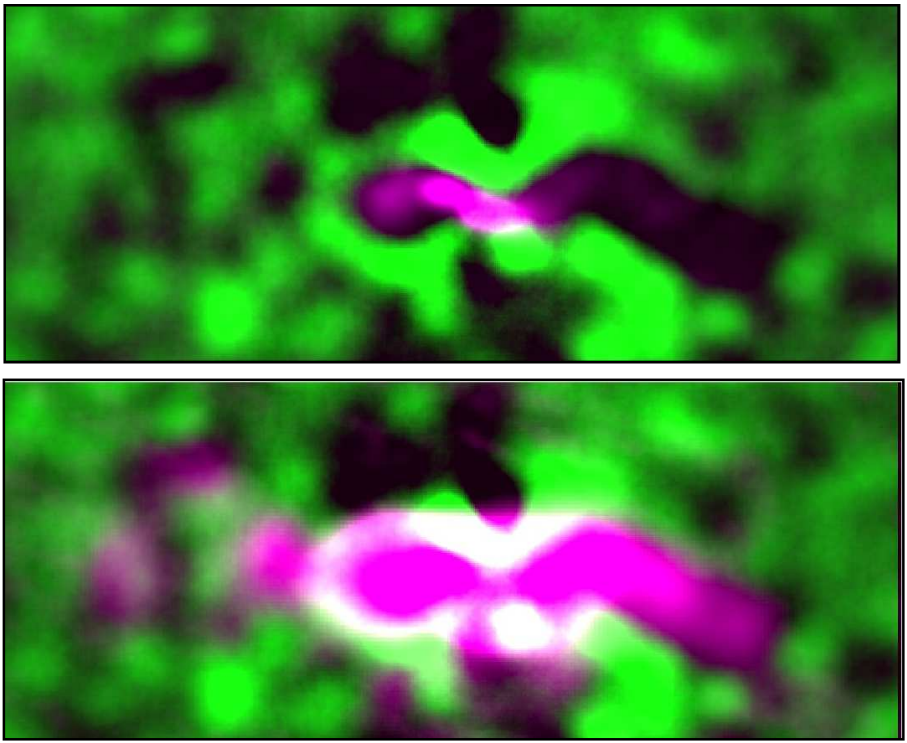}
\caption{Residual {\it Chandra} 139 ksec $3'.7\times1'.5$ ($70\times30$ kpc) 
images (green) of Abell 262 with radio emission superposed in 
pink (VLA 1.4 GHz [top] and GMRT 610 MHz [bottom], \cite{clarke09}).}
\end{figure}

\end{document}